\documentclass{SciPost}

\definecolor{bgclr}{RGB}{240, 240, 240}
\definecolor{commentclr}{RGB}{0, 136, 0} 
\definecolor{keywordclr}{RGB}{170, 34, 255} 
\definecolor{packageclr}{RGB}{0, 0, 255}
\definecolor{stringclr}{RGB}{187, 34, 34}
\usepackage{listings}
\lstdefinestyle{pythonstyle}{
    language=Python,
    numbers=left,
    morekeywords=[1]{as},
    frame=single,
    framesep=3pt,
    backgroundcolor=\color{bgclr},
    basicstyle={\ttfamily\scriptsize},
    keywordstyle=[1]\color{keywordclr},
    keywordstyle=[2]\color{keywordclr},
    keywordstyle=[3]\color{packageclr},
    commentstyle=\color{commentclr},
    stringstyle=\color{stringclr},
    breaklines=true,
    tabsize=4,
    captionpos=b
}
\newcommand{\narrowtt}[1]{%
    \scalebox{0.9}[1.0]{\texttt{#1}}%
}

\hypersetup{
    colorlinks,
    linkcolor={red!50!black},
    citecolor={blue!50!black},
    urlcolor={blue!80!black}
}

\usepackage[bitstream-charter]{mathdesign}
\usepackage{orcidlink}
\usepackage{siunitx}
\DeclareSymbolFont{usualmathcal}{OMS}{cmsy}{m}{n}
\DeclareSymbolFontAlphabet{\mathcal}{usualmathcal}

\fancypagestyle{SPstyle}{
\fancyhf{}
\lhead{}
\rhead{}

\fancyfoot[C]{\textbf{\thepage}}
}
\usepackage{url}
\usepackage{hyperref}

\usepackage{dsfont}
\usepackage{siunitx}

\usepackage{makecell}
\usepackage[normalem]{ulem}

\begin{document}
\pagestyle{SPstyle}

\begin{center}{\Large \textbf{
Q-GAIN: A Python Package for Machine Learning and Physically Informed Analysis Applications\\
}}\end{center}

\begin{center}\textbf{
Michael Doris\orcidlink{0009-0000-5844-8148}\textsuperscript{1,2$\star$},
Shangjie Guo\orcidlink{0000-0002-6910-636X}\textsuperscript{1,2},
Sophia M. Koh\orcidlink{0000-0001-5481-0285}\textsuperscript{1,3},
Lisa Ritter\orcidlink{0000-0003-1334-3304}\textsuperscript{1},
Amilson~R.~Fritsch\orcidlink{0000-0003-1419-8591}\textsuperscript{1,2},
Shouvik Mukherjee\orcidlink{0000-0003-0217-0743}\textsuperscript{2},
I.~B.~Spielman\orcidlink{0000-0003-1421-8652}\textsuperscript{1,2},
Justyna P. Zwolak\orcidlink{0000-0002-2286-3208}\textsuperscript{1,4$\dagger$}
}\end{center}

\begin{center}
{\bf 1} National Institute of Standards and Technology, Gaithersburg, MD 20899, USA
\\
{\bf 2} Joint Quantum Institute and University of Maryland, College Park, MD 20742, USA
\\
{\bf 3} Department of Physics and Astronomy, Amherst College, Amherst, MA 01002, USA
\\
{\bf 4} Joint Center for Quantum Information and Computer Science,\\University of Maryland, College Park, MD 20899, USA
\\[\baselineskip]
$\star$ \href{mailto:mjdoris@terpmail.umd.edu}{\small mjdoris@umd.edu}\,,\quad
$\dagger$ \href{mailto:jpzwolak@nist.gov}{\small jpzwolak@nist.gov}
\end{center}
\section*{Abstract}
\textbf{\boldmath{%
Here we describe the quantum gas analysis and inference (\texttt{Q-GAIN}) Python package, which enables rapid deployment of machine learning (ML) and physics-informed analysis techniques for cold-atom experiments.
Out of the box, \texttt{Q-GAIN} implements classification, object detection, and physics-informed metrics for feature detection in images of atomic Bose-Einstein condensates (BECs).
\texttt{Q-GAIN} encourages a natural, module-based workflow: starting with data loading and preprocessing, followed by ML-based feature identification, and ending with conventional analysis techniques.
We demonstrate this modularity by configuring \texttt{Q-GAIN} for three ML tasks.
First, we demonstrate the basic workflow of the \texttt{Q-GAIN} framework by implementing the standard task of classifying handwritten digits from the MNIST dataset.
Then, we re-implement our earlier soliton detection (SolDet) package in the \texttt{Q-GAIN} framework, enabling the detection and analysis of solitonic excitations in time-of-flight data. 
Finally, we develop an object-detection tool that identifies quantized vortices in images of ring-shaped BECs.
}}
\vspace{\baselineskip}

\section{Introduction}
\label{sec:intro}
Machine learning (ML) provides powerful tools for exploring, organizing, and analyzing scientific data.
Applications range from automated data curation~\cite{armstrong19-FSC} and feature identification~\cite{carrasquilla17-MPM, stanev18-MLT} to dimensionality reduction and visualization of high-dimensional datasets~\cite{maaten08-SNE, mardt18-MLM}.
In ultracold atom experiments, many ML tasks follow a common workflow involving data preprocessing, feature extraction, classification, and statistical interpretation.
However, analysis pipelines are often developed independently for individual studies or adapted from pretrained architectures in ways that are tightly coupled to specific datasets or experimental systems~\cite{ziatdinov22-AAI}.
As a result, substantial effort is frequently required to reproduce, modify, or extend existing implementations for new applications.
This presents an opportunity to streamline the process by providing a unified framework that simplifies the development and deployment of these pipelines.

Here, we present \textit{\textbf{Q}uantum \textbf{G}as \textbf{A}nalysis and \textbf{I}nference} (\texttt{Q-GAIN}), a Python framework for developing and deploying ML- and statistics-based analysis tools for scientific imaging data, with a specific focus on cold-atom quantum gas experiments.
\texttt{Q-GAIN} provides an object-oriented, modular interface that supports reusable analysis workflows while encouraging the adaptation, replacement, and extension of its built-in components.
The framework integrates tools for retrieving and preprocessing experimental data, maintaining compatibility with existing analysis pipelines, ML-based object classification and tracking, and physics-informed statistical analysis.
Its modular architecture allows existing analysis modules to be applied directly to new labeled or unlabeled datasets, or adapted to systems with different physical parameters and imaging conditions.

Although \texttt{Q-GAIN} was originally developed for ultracold atom systems, the framework is not restricted to this domain.
New analysis environments can be supported by implementing specialized modules while retaining the layered architecture for data handling, model deployment, and statistical analysis.
This design simplifies the development of new analysis workflows by separating domain-specific functionality from shared infrastructure.
The framework supports workflows spanning data preprocessing, ML-assisted feature tagging and transformation, downstream statistical analysis, and export of processed results into multiple file formats for dataset generation and curation.

We first present a high-level overview of the \texttt{Q-GAIN} framework using a standard ML task: the classification of handwritten digits from the MNIST dataset~\cite{lecun10-MHD} (in Section~\ref{sec:mnist}).
This example demonstrates how to construct new specialized classes and how to repurpose existing models and tools.
In Section~\ref{ssec:soldet}, we discuss a reimplementation within \texttt{Q-GAIN} of an existing Python library, SolDet~\cite{guo22_SSF}. 
Finally, in Section~\ref{ssec:vdet}, we present a new specialized \scalebox{0.8}[1.0]{\narrowtt{VortexDetector}} class for identifying and counting vortices in images of ring-shaped Bose-Einstein condensates (BECs), using ML architectures inspired by Ref.~\cite{metz21-DLV}.

\texttt{Q-GAIN} is available for download from the \href{https://github.com/Q-GAIN}{Q-GAIN GitHub organization}, with source code distributed under the GNU Lesser General Public License in the \texttt{Q-GAIN} GitHub repository~\cite{doris25_QGN}.
Comprehensive API documentation is provided via docstrings embedded in the source code and as HTML pages hosted on \href{https://q-gain.github.io}{GitHub Pages}.
The library includes type hints for all code to support clarity, maintainability, and extensibility.
Unit tests are also distributed with the \texttt{Q-GAIN} repository to facilitate validation and performance evaluation.

\section{Framework overview}
\label{sec:library}
\begin{figure}[!t]
    \centering
    \includegraphics[width=0.8\linewidth]{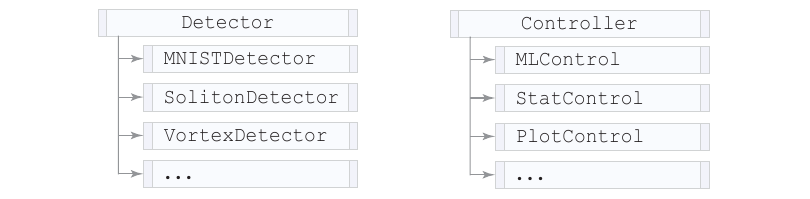}
    \caption{Examples of implementations of the two main \texttt{Q-GAIN} classes.
    The \narrowtt{Detector} class (left) provides the core functionality to each subclass, while the \narrowtt{Controller} class (right) provides specialized capabilities that are shared between \narrowtt{detector\_objects}.
    }
    \label{fig:overview}
\end{figure}

The core functionality of \texttt{Q-GAIN} is built around a base \narrowtt{Detector} class, which provides a unified interface for constructing reusable analysis workflows.
Instances of the \narrowtt{Detector} class, or any of its subclasses, are referred to as \narrowtt{detector\_object}s.
These objects provide access to the framework's primary capabilities, including data handling, ML model training and inference, statistical analysis, and visualization.
Their behavior can be customized through configurable modules and user-defined function calls supplied during initialization.
A simplified overview of the \texttt{Q-GAIN} class structure is shown in Fig.~\ref{fig:overview}(a).

\texttt{Q-GAIN} provides flexible data-handling functionality for importing and preprocessing experimental data, loading it into a detector, and exporting analysis results in various formats. 
By default, \narrowtt{detector\_object}s support importing absorption images from ultracold atom experiments stored in HDF files generated by Labscript~\cite{starkey13-SCS} and provide preprocessing routines, including image cropping, masking, and optical-depth calculation.
Support for additional data modalities, including alternative image formats, audio, or text data, can be added by supplying custom preprocessing routines during initialization.
Metadata and labels associated with labeled datasets can also be incorporated during the import stage.

Specialized analysis functionality within \texttt{Q-GAIN} is implemented via modular components called ``tools.''
These tools are managed through subclasses of the \narrowtt{Controller} class, which provides a shared interface for deploying analysis routines.
The \narrowtt{Controller} subclasses are specialized for distinct categories of analysis tasks while maintaining a consistent code infrastructure.
Users can extend the framework by implementing new controllers or integrating additional tools into existing ones.
At present, \texttt{Q-GAIN} includes three primary controller subclasses: \narrowtt{MLControl}, \narrowtt{StatControl}, and \narrowtt{PlotControl}.

The \narrowtt{MLControl} class provides ML-related functionality, including training and inference of PyTorch-compatible models.
Currently, \texttt{Q-GAIN} supports classification and object-detection workflows through built-in ML tools.
Additional ML models can be directly integrated into \narrowtt{MLControl} using the \narrowtt{add\_new\_tool()} method.
Users can configure standard training parameters such as batch size, number of training epochs, learning rates, optimizers, early stopping criteria, and model checkpointing.
Saved model weights can subsequently be reloaded to initialize pretrained models for inference or continued training.
The framework also supports configurable validation metrics that are tracked and stored during training.

The statistical analysis capabilities are provided by the \narrowtt{StatControl} class, which manages deployable statistical analysis tools, referred to as \narrowtt{stat\_tool}s.
A \narrowtt{stat\_tool} implements downstream analysis routines that operate on ML-generated or preprocessed data products.
New statistical analysis methods can be integrated into \narrowtt{StatControl} using the same modular interface as \narrowtt{MLControl}.
Configuration parameters and fitted model parameters can optionally be saved and reloaded to support reproducible analysis workflows.

Visualization functionality is implemented through the \narrowtt{PlotControl} class, which manages plotting-oriented tools referred to as \narrowtt{plot\_tool}s.
By default, \texttt{Q-GAIN} includes plotting utilities for visualization tasks such as object-detection scatter plots and classifier confusion matrices.
Additional plotting routines can be integrated through the \narrowtt{add\_new\_tool()} method.
The framework supports rendering through Matplotlib~\cite{hunter07-PLT}, including optional use of style sheets and automated export of generated figures.

A typical \texttt{Q-GAIN} workflow consists of importing and preprocessing experimental data, applying ML-based feature identification or classification, and performing downstream statistical analysis.
Many ML and statistical analysis tools require an initial training or fitting stage before deployment on new systems or datasets.
Once trained, these tools can be reused to automate analysis of new data, with results exported in multiple file formats or visualized directly via the framework's plotting utilities.

\section{Minimal example: MNIST dataset}
\label{sec:mnist}
To demonstrate the modular design and extensibility of \texttt{Q-GAIN}, we implement a handwritten digit classifier using the MNIST dataset~\cite{lecun10-MHD}.
MNIST is a standard ML benchmark rather than a physics application, and it serves as a compact example of how \texttt{Q-GAIN} separates data management, model execution, statistical analysis, and visualization into reusable, interchangeable components.

The architecture of \texttt{Q-GAIN} is designed to reduce coupling between analysis stages while enabling workflows developed for one application to be adapted to new datasets and experimental systems with minimal task-specific code.
The MNIST example demonstrates this design philosophy by combining custom preprocessing routines with the ML and statistical analysis infrastructure already available within the framework.

In this example, a specialized \narrowtt{MNISTDetector} subclass is constructed by extending the base \narrowtt{Detector} interface with dataset-specific preprocessing routines and lightweight wrappers around existing analysis tools.
The workflow consists of importing and preprocessing raw data, loading processed samples into the detector, training and deploying ML models, and applying downstream statistical-analysis routines.

\begin{figure}[!t]
    \centering
    \includegraphics[]{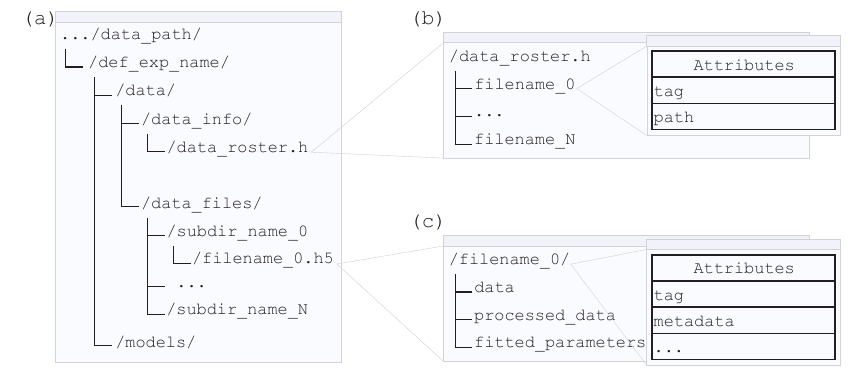}
    \caption{The directory and file structure for \texttt{Q-GAIN}.
    (a) Each experimental directory contains a \narrowtt{data} subdirectory with the data roster stored in \narrowtt{data\_info} and data files stored in \narrowtt{data\_files}, and a \narrowtt{models} subdirectory containing trained ML models and fitting files.
    All exported data is also stored in the experiment directories.
    (b) The basic structure of the HDF roster file system. 
    Each entry points to its corresponding file on disk and contains a tag attribute that provides a descriptive value to distinguish one file from another. 
    Each HDF data file, shown in (c), contains a root group whose attributes hold all required and optional metadata provided by the user. 
    Datasets within an HDF group hold the target data as well as any additional dictionaries or arrays of information.}
    \label{fig:structure}
\end{figure}

\subsection{Data management and preprocessing}
\label{ssec:data_handling}
\texttt{Q-GAIN} organizes analysis tasks using a directory-based experiment structure that separates raw and processed data from trained models, statistical analysis objects, and exported results.
Each experiment directory contains the data products associated with a specific analysis workflow, along with the metadata required to reproduce the workflow.
A schematic overview of the resulting data organization within \texttt{Q-GAIN} is shown in Fig.~\ref{fig:structure}.

Within an experiment directory, \texttt{Q-GAIN} maintains a roster file that tracks the locations and descriptors of all imported data samples.
Each entry in the roster corresponds to a single data point and contains metadata used to organize and retrieve the associated files.
These descriptors are represented using user-defined tags that can encode classification labels, dataset partitions, or other application-specific identifiers.
The roster system allows labeled and unlabeled datasets to be managed using a common interface while enabling data subsets to be selectively loaded during later analysis stages.

Data are imported into an experiment through the detector's \narrowtt{import\_data()} method.
During import, a user-defined preprocessing function transforms raw input data into a format compatible with \texttt{Q-GAIN}.
This preprocessing stage also provides a mechanism for attaching metadata and labels to individual samples.
To support extensibility across different data modalities, preprocessing functions are supplied during detector initialization through the \narrowtt{process\_fn} argument.

\texttt{Q-GAIN} expects preprocessing functions to return a list-like object containing dictionaries with, at minimum, the keys \narrowtt{data}, \narrowtt{tag}, and \narrowtt{sub\_dir}.
The \narrowtt{data} field stores the target data used during analysis, while \narrowtt{sub\_dir} specifies the experiment subdirectory in which the sample should be stored.
The \narrowtt{tag} field provides a descriptive identifier that can subsequently be used for dataset filtering, partitioning, or as a supervised-learning target.
Additional metadata can be included through arbitrary user-defined entries.
If no preprocessing function is supplied, \texttt{Q-GAIN} defaults to built-in routines for importing absorption images from Labscript~\cite{starkey13-SCS} experiment files.

\noindent
\begin{figure}[!ht]
\begin{lstlisting}[
caption={Example code for importing new data into \texttt{Q-GAIN} to support the classification of MNIST data.
\texttt{Q-GAIN} expects a preprocessing function that returns a list-like object of dictionaries.
These dictionaries can contain any number of descriptive metadata entries, but at a minimum should contain the keys \narrowtt{data}, \narrowtt{tag}, and \narrowtt{sub\_dir}.},
label={lst:mnist_import}]
class MNISTDetector(Detector):
    def __init__(self) -> None:
        super().__init__(process_fn=self.mnist_process_fn)

    def mnist_process_fn(self, input_data: MNIST, *, train: bool = True) -> list[dict]:
        data_samples = []
        data_set = MNIST(input_data, train=train, download=True)
        for (image, label) in tqdm(data_set, desc="Processing data"):
            sample = {}
            sample["data"] = image
            sample["tag"] = "training" if train else "testing"
            sample["sub_dir"] = sample["tag"]
            sample["label"] = label
            data_samples += [sample]
        return data_samples

mnist_det = MNISTDetector()
mnist_det.import_data(exp_path, **{"train": True})
mnist_det.import_data(exp_path, **{"train": False})
mnist_det.load_data(tags="training", minmax=[0, 255], scale=True)
\end{lstlisting}
\end{figure}

During import, processed samples and associated metadata are serialized into hierarchical data format (HDF) files.
A simplified representation of the roster and data-file organization is shown in Fig.~\ref{fig:structure}(b) and Fig.~\ref{fig:structure}(c), respectively.
The framework stores datasets along with user-provided metadata, ensuring that experimental parameters, labels, and auxiliary analysis quantities remain associated with individual data samples throughout the workflow.

Listing~\ref{lst:mnist_import} demonstrates the preprocessing pipeline for the MNIST example.
A custom preprocessing function downloads the MNIST dataset and packages the resulting arrays together with dataset-specific metadata.
The \narrowtt{tag} and \narrowtt{sub\_dir} entries are used to separate training and testing data into distinct experiment subdirectories, while the handwritten digit labels are stored through an additional \narrowtt{label} metadata entry.
The preprocessing function is supplied during initialization of the \narrowtt{MNISTDetector}, after which the data can be imported with calls to \narrowtt{import\_data()}.

Once imported, datasets can be reloaded into a detector object repeatedly without additional preprocessing.
The \narrowtt{load\_data()} method retrieves samples associated with selected tags and optionally applies normalization operations to array-like data, as shown in line 21 in Listing~\ref{lst:mnist_import}.
This design separates the computationally expensive import and preprocessing stages from later training and inference workflows, enabling datasets to be reused across multiple analysis tasks.

\subsection{ML controller and model integration}
\label{ssec:MLControl}
ML functionality within \texttt{Q-GAIN} is managed through the \narrowtt{MLControl} controller class, which provides a unified interface for training, inference, and deployment of PyTorch-compatible models.
The controller encapsulates individual ML analysis routines within modular components referred to as \narrowtt{MLTool}s.
This abstraction allows heterogeneous models and datasets to share a common workflow interface while remaining independent of specific ML architectures or data representations.

\narrowtt{MLTool}s are typically initialized during detector construction, but can also be added dynamically through the \narrowtt{add\_new\_tool()} method.
Each tool is defined through three primary components: a PyTorch model, a dataset interface compatible with PyTorch \narrowtt{DataLoader}s, and a loss function.
Because \texttt{Q-GAIN} imposes no constraints beyond standard PyTorch interfaces, existing neural-network architectures and dataset classes can be integrated directly into the framework with minimal modification.
This design enables previously developed models to be reused across new analysis workflows while retaining a consistent training and inference interface.

The MNIST example demonstrates this interoperability by specializing one of the built-in \texttt{Q-GAIN} architectures, rather than introducing a new network model.
This generalized classification model operates by employing convolutional layers and residual connections~\cite{ziegler23_AQD}.
To adapt the classifier to handwritten digit recognition, the number of output classes is modified through the initialization parameter \narrowtt{num\_classes}.
The number of convolution stacks and their kernel sizes are configured with the \narrowtt{stack\_size} and \narrowtt{kernel\_size} parameters, respectively.
The number of layers is set by specifying a list of filter sizes with the parameter \narrowtt{filter\_list}.
The remainder of the training and inference workflow is handled directly by \texttt{Q-GAIN}.
A lightweight dataset wrapper is additionally implemented to expose the imported MNIST samples through the standard PyTorch dataset interface.
The resulting implementation is shown in Listing~\ref{lst:mnist_ML}.

\noindent
\begin{figure}[!ht]
\begin{lstlisting}[
caption={Example code for setting up the MNIST detector's ML controller.
Rather than writing a new model, the classifier from the SolDet sub-package is used.
A simple Dataset class is written to return images and their corresponding labels.
A standard loss function from the PyTorch library is also provided.
The ML controller can be instructed to train the classifier by using the {\tt train\_nn()} method.
To run inference, the {\tt use\_models()} method is overridden to convert the classifier's output to class values.},
label={lst:mnist_ML}]]
class MNISTDetector(Detector):
    def __init__(self) -> None:
        super().__init__(cl_model=qgain.models.MLST2021CNNmodern,
                         cl_kwargs={"filter_list": [8, 16, 32, 64], "stack_size": 1, "kernel_size": 4, "num_classes": 10},
                         cl_dataset_fn=self.MNISTDataset,
                         cl_aug=None,
                         cl_loss_fn=torch.nn.NLLLoss)

    class MNISTDataset(torch.utils.data.Dataset):
        def __init__(self, data: list[dict] | tuple[dict]) -> None:
            self.data = list(data)

        def __len__(self) -> int:
            return len(self.data)

        def __getitem__(self, idx: int) -> tuple[torch.Tensor, int]:
            img = self.data[idx]["data"]
            img = torch.from_numpy(img[np.newaxis, np.newaxis, :]).float()
            label = int(self.data[idx]["label"]) if "label" in self.data[idx] else None
            return img, label

    def use_models(self, model_paths: list, model_list: list | tuple,
                   data: list | dict | None = None) -> None:
        super().use_models(model_list=model_list, model_paths=model_paths, data=data)
        target_data = self.data if data is None else data
        for item in target_data:
            if "CL_pred" in item:
                item["CL_pred"] = np.argmax(item["CL_pred"])
        if data is None:
            self.data = target_data

mnist_det = MNISTDetector()
mnist_det.load_data(tags="training", data_frac=0.9, minmax=[0, 255], scale=True)
mnist_det.train_nn(epochs=100, patience=10)
mnist_det.load_data(tags=["testing"], minmax=[0, 255], scale=True, keep=False)
mnist_det.use_models(model_list=["CL"], model_paths=["20260223_190725_CL.pt"])
\end{lstlisting}
\end{figure}

\texttt{Q-GAIN} supports configurable training workflows with the detector method \narrowtt{train\_nn()}, which exposes common optimization parameters including learning rate, batch size, optimizer selection, training epochs, and early stopping criteria; see Table~\ref{tab:train_params}.
Models can be trained using either internally loaded detector data or externally supplied datasets.
Trained model weights may optionally be saved and later reloaded for inference or continued optimization.

\begin{table}[!t]
    \centering
    \begin{tabular}{l l l}
        \hline\hline
        \multicolumn{1}{c}{\textbf{Parameter}} & 
        \multicolumn{1}{c}{\textbf{Description}} & 
        \multicolumn{1}{c}{\textbf{Type}}  \\
        \hline
        \narrowtt{optimizer\_fn} & The optimizing function to use during training. & PyTorch Optimizer\\
        \narrowtt{model\_path} & The path to where weights should be saved. & String or None\\
        \narrowtt{batch\_size} & The size of the batch a dataset is divided into. & Integer \\
        \narrowtt{epochs} & The number of cycles through the entire dataset. & Integer \\
        \narrowtt{patience} & The cycle count to wait with no improvement. & Integer \\
        \narrowtt{lr} & The learning rate used by the optimizer. & Float \\ 
        \hline\hline
    \end{tabular}
    \caption{A brief overview of selected training parameters.}
    \label{tab:train_params}
\end{table}

Inference is performed using the \narrowtt{use\_models()} method, which applies trained \narrowtt{MLTool}s to loaded data samples.
For the MNIST example, the inherited \narrowtt{use\_models()} method is lightly overridden to convert the probability outputs of the classifier into discrete digit labels.
This modification illustrates how detector subclasses can adapt existing analysis models to application-specific output formats while preserving the underlying framework workflow.

\noindent
\begin{figure}[!ht]
\begin{lstlisting}[
caption={Example code demonstrating the configuration of optional runtime parameters.
Changing the device on which a neural network model runs is done by setting the \narrowtt{device} attribute.
It is also possible to add metrics reported during training.
This is achieved by setting the \narrowtt{metrics} attribute, which expects a list of dictionaries containing the metric name and the function call that computes the loss value.},
label={lst:ML_params}]
class MNISTDetector(Detector):
    def __init__(self) -> None:
        super().__init__(process_fn=self.mnist_process_fn,
                         cl_model=qgain.models.MLST2021CNNmodern,
                         cl_kwargs={"filter_list": [8, 16, 32, 64], "stack_size": 1, "kernel_size": 4, "num_classes": 10},
                         cl_dataset_fn=self.MNISTDataset,
                         cl_aug=None,
                         cl_loss_fn=torch.nn.NLLLoss)
        
        self.controllers["ML Controller"].add_new_tool(model=DummyModel, name="DummyModel", dataset_fn=DummyDataset, loss_fn=torch.nn.SmoothL1Loss, device=0, metrics=[{"name": "SmoothL1", "metric": torch.nn.SmoothL1Loss()}])

        self.controllers["ML Controller"].get_tool("CL").device = "cuda:1"
        self.controllers["ML Controller"].get_tool("CL").metrics += [{"name": "GaussianNL", "metric": torch.nn.GaussianNLLLoss()}]   
\end{lstlisting}
\end{figure}

Beyond model execution, \narrowtt{MLControl} supports additional runtime configuration, including device selection and user-defined validation metrics.
Models can be deployed on CPUs or CUDA-enabled GPUs, and arbitrary validation metrics may be incorporated alongside built-in loss tracking during training.
These capabilities are exposed through configurable \narrowtt{MLTool} attributes and are illustrated in Listing~\ref{lst:ML_params}.
Together, these components provide a reusable infrastructure for integrating ML models into scientific-analysis workflows while separating model-specific implementation details from workflow orchestration and data management.

\subsection{Statistical analysis tools}
\label{ssec:StatsControl} 
In addition to ML workflows, \texttt{Q-GAIN} provides infrastructure for integrating downstream statistical analysis routines via the \narrowtt{StatControl} controller class.
Statistical analysis methods are encapsulated in modular components, referred to as \narrowtt{stat\_tool}s, and are managed using the same controller-based architecture as ML workflows.
This design allows statistical inference routines to operate directly on imported, preprocessed, or ML-generated data products while remaining integrated within a unified analysis pipeline.

New \narrowtt{stat\_tool}s can be attached to a detector during initialization or dynamically added through the \narrowtt{add\_new\_tool()} method.
Each tool is implemented as a lightweight wrapper around a statistical-analysis algorithm and is expected to expose a \narrowtt{transform()} method that operates on collections of detector data entries.
Optionally, \narrowtt{stat\_tool}s may implement a \narrowtt{fit()} method for training or parameter estimation before deployment.
This interface allows external analysis packages to be incorporated into \texttt{Q-GAIN} while preserving a consistent workflow abstraction.

The MNIST example demonstrates this capability using a K-Means clustering implementation based on Scikit-learn~\cite{Pedregosa11_SKL}.
A custom \narrowtt{KMeansTool} wrapper is introduced to adapt the Scikit-learn implementation to the \texttt{Q-GAIN} data interface.
The wrapper's \narrowtt{fit()} method extracts image arrays from detector data entries, performs clustering, and constructs a mapping between cluster assignments and handwritten digit labels.
The corresponding \narrowtt{transform()} method subsequently applies the trained clustering model to new samples and converts the predicted cluster indices into digit classifications.
The resulting implementation is shown in Listing~\ref{lst:mnist_stat}.

\noindent
\begin{figure}[!ht]
\begin{lstlisting}[
caption={Example code for setting up a \narrowtt{stat\_tool}.
A \narrowtt{fit()} method is written to retrieve the data arrays, call the K-Means fit function, and create a mapping between clusters and digit labels.
A \narrowtt{transform()} method is written to call the K-Means prediction algorithm and return the predicted digit label.
The tool is added to the MNIST detector by using its initialization arguments.
Data is loaded separately for fitting and prediction using the built-in methods.},
label={lst:mnist_stat}]
class MNISTDetector(Detector):
    def __init__(self) -> None:
        super().__init__(stat_tools=[{"name": "KMeans", "tool": self.KMeansTool},], stats_kwargs=[{"n_clusters": 10,}])
    
    class KMeansTool(KMeans):
        def __init__(self, **kwargs: dict):
            super().__init__(**kwargs)
        
        def fit(self, data: list[dict]):
            target_data = []
            self.clust_to_labels = np.zeros((10))
            gmatrix = np.zeros((10,10))
            for sample in data:
                target_data += [sample["data"].flatten()]
            print("Fitting to data.")
            super().fit(np.array(target_data))
            for ground, pred in zip(data, self.labels_):
                gmatrix[int(pred), int(ground["label"])] += 1
            for cluster in range(10):
                self.clust_to_labels[cluster] = gmatrix[cluster, :].argmax()
            del gmatrix
            del target_data

        def transform(self, data: list[dict]):
            target_data = []
            predictions = []
            for sample in data:
                target_data += [sample["data"].flatten()]
            print("Determining predictions.")
            result = self.predict(np.array(target_data))
            for res in result:
                predictions += [self.clust_to_labels[res]]
            del result
            del target_data
            return predictions

mnist_det = MNISTDetector()
mnist_det.load_data(tags=["training"], minmax=[0, 255], scale=True)
mnist_det.define_stat(save=True)
mnist_det.load_data(tags=["testing"], minmax=[0, 255], scale=True, keep=False)
mnist_det.use_models(model_list=["KMeans"], model_paths=["20260223_202454_KMeans.pkl"])
\end{lstlisting}
\end{figure}

Statistical analysis workflows are deployed via the same detector-level interface used for ML inference.
The \narrowtt{define\_stat()} method initializes and fits statistical-analysis tools when required, while \narrowtt{use\_models()} applies one or more fitted tools to loaded detector data.
Because fitted analysis objects can optionally be serialized and reloaded, computationally expensive fitting procedures need not be repeated for subsequent analysis tasks.

Together, the \narrowtt{StatControl} infrastructure and the \narrowtt{stat\_tool} abstraction enable classical statistical analysis methods and ML workflows to coexist within a common scientific analysis framework.
This allows downstream inference stages to be composed from heterogeneous analysis methods while maintaining a unified data-management and execution interface.

\subsection{Export and reproducibility infrastructure}
\label{ssec:export}
\texttt{Q-GAIN} includes infrastructure for saving and reusing analysis results, trained models, fitted statistical-analysis tools, and processed datasets to support reproducible workflows.
Because imported data, preprocessing outputs, trained model weights, and fitted analysis objects are stored within the experiment directory structure, analysis pipelines can be resumed or reapplied without repeating earlier computational stages.

Processed detector data and analysis results can be exported through the detector method \narrowtt{export()} using multiple output formats, including comma-separated values, HDF, HTML, pickle, and NumPy object arrays.
Exported quantities are selected by specifying the corresponding detector-data keys, allowing subsets of analysis outputs to be shared or integrated into external workflows.
This functionality enables \texttt{Q-GAIN} to interoperate with downstream analysis environments while maintaining a consistent internal data representation.

The framework also supports generating reusable datasets via the \narrowtt{generate\_samples()} method.
This method creates portable copies of experiment data structures containing user-selected data fields and metadata entries.
The resulting datasets can subsequently be distributed, re-imported into independent \texttt{Q-GAIN} experiments, or reused in alternative training and analysis workflows without requiring repeated preprocessing of the original raw data.

The reproducibility infrastructure of \texttt{Q-GAIN} extends beyond dataset management to include the persistence of trained ML models and fitted statistical analysis tools.
Saved model weights can be reloaded for inference or continued training, while fitted \narrowtt{stat\_tool} objects can be serialized and restored without repeating computationally expensive fitting procedures.
Together, these capabilities allow complete analysis pipelines to be reconstructed from stored experiment directories, facilitating iterative development and reproducible scientific analysis.

\section{Scientific applications}
\label{sec:applications}
The primary goal of \texttt{Q-GAIN} is to provide a reusable framework for combining ML and statistical analysis workflows within a consistent scientific analysis environment.
In this section, we present two applications that demonstrate how this architecture can support both the implementation of existing analysis pipelines and the development of new domain-specific workflows.

Section~\ref{ssec:soldet} presents the integration of the SolDet soliton detector analysis framework into \texttt{Q-GAIN}, illustrating how an established scientific ML workflow can be migrated into the framework while preserving domain-specific functionality and reusing shared infrastructure.
Section~\ref{ssec:vdet} then demonstrates the extensibility of the framework through the implementation of a specialized vortex-detection workflow for ring-shaped BECs.
Together, these examples illustrate how \texttt{Q-GAIN} separates reusable workflow infrastructure from application-specific scientific analysis logic.

\subsection{Soliton detection and classification}
\label{ssec:soldet}
The SolDet sub-package illustrates how an existing scientific analysis workflow can be incorporated into \texttt{Q-GAIN}.
Originally developed for automated identification and characterization of solitons in BECs, SolDet combines ML-based excitation detection with physically informed statistical analysis methods.
Its integration into \texttt{Q-GAIN} demonstrates how specialized scientific workflows can be modernized, generalized, and extended while reusing shared infrastructure for data management, model execution, and reproducible analysis.

\subsubsection{Integration of SolDet into Q-GAIN}
SolDet was originally developed as an ML framework for automated localization and identification of solitonic excitations in BECs~\cite{guo21-MLD}.
The original implementation combined ML-based detection and classification with physically informed statistical analysis to characterize identified excitations.
However, as the project evolved, maintaining the standalone implementation became increasingly challenging due to growing software-library incompatibilities and workflow constraints tied to experiment-specific data formats.

To improve maintainability, extensibility, and interoperability, SolDet's codebase was updated and integrated into \texttt{Q-GAIN} as a dedicated sub-package.
Within this framework, SolDet subclasses the base \narrowtt{Detector} class to construct the specialized \narrowtt{SolitonDetector} class.
This migration preserved the original scientific analysis workflow while replacing custom infrastructure for data management, preprocessing, model execution, and statistical analysis orchestration with reusable components provided by \texttt{Q-GAIN}.

Integration into \texttt{Q-GAIN} also generalized SolDet's data-management workflow.
The original implementation relied on a highly specific HDF file organization scheme associated with the experimental system on which it was initially developed.
By adopting the modular preprocessing and experiment infrastructure of \texttt{Q-GAIN}, the modernized SolDet implementation can now support alternative HDF organization schemes and experimental datasets through configurable preprocessing routines.

Finally, the migration enabled a PyTorch-based ML infrastructure.
The original neural network implementations were replaced with specialized classification and object-detection architectures provided by \texttt{Q-GAIN}.
This enabled improved interoperability with modern ML tooling while preserving compatibility with the broader \texttt{Q-GAIN} workflow.
This redesign additionally introduced architectural refinements, including residual connections and deeper convolutional structures, as well as improvements to the physically informed statistical analysis routines to better support images containing multiple excitations.

The resulting \narrowtt{SolitonDetector} demonstrates how an existing scientific-analysis workflow can be incorporated into \texttt{Q-GAIN} while preserving domain-specific functionality and reusing shared framework infrastructure.

\subsubsection{ML-based soliton detection and classification}
The \narrowtt{SolitonDetector} integrates two coordinated ML models for automated analysis of BEC images: an object detector for localization of solitonic excitations and a classifier for identifying excitation classes.
Together, these models provide the initial analysis stage of the SolDet workflow before downstream physics-informed statistical analysis is applied.

The classifier assigns the image into one of three classes defined based on the number of potential excitations (\narrowtt{class-0}: no excitations; \narrowtt{class-1}: a lone excitation; and \narrowtt{class-2}: ambiguous or multiple excitations), while the object detector estimates the excitation positions within the BEC image.
To support localization across images with different spatial resolutions and feature sizes, the object detector operates in a compressed cell-space representation, where each cell encodes both the probability that a soliton is present and its approximate position in the original image.
The size of this representation can be configured during detector initialization to accommodate different imaging conditions and experimental systems.

Both ML models are integrated into \texttt{Q-GAIN} through the standard \narrowtt{MLControl} interface described in Section~\ref{ssec:MLControl}.
The \narrowtt{SolitonDetector} supplies custom dataset wrappers for both the classifier and object detector, enabling detector data entries to be converted into PyTorch-compatible datasets while preserving associated metadata and labels.
The dataset wrappers additionally support optional data augmentation through horizontal and vertical reflections and $180^\circ$ rotations.

The classifier model uses a negative log-likelihood loss function, while the object detector employs a custom loss function based on a one-dimensional adaptation of the localization framework introduced in Ref.~\cite{metz21-DLV}.
This loss combines a cross-entropy term that determines the presence of solitons within individual cells with a mean-squared-error term that refines their predicted positions.
Additional validation metrics, including classification and detection accuracy, are incorporated through the configurable metric infrastructure provided by \texttt{Q-GAIN}.

\noindent
\begin{figure}[!ht]
\begin{lstlisting}[
caption={A simplified overview of how the \narrowtt{SolitonDetector} plugs in its modules to set up a detector for soliton tracking, classification, and analysis.
The initialization arguments are used to plug in the ML and statistical-based tools.
The {\tt use\_models()} method is overridden to convert the ML model results to the expected output format.
Full code and documentation are available in the GitHub repository~\cite{doris25_QGN}.},
label={lst:soldet}]
class SolitonDetector(Detector):
    def __init__(self, od_kwargs: dict | None = None, cl_kwargs: dict | None = None,
                 *, augment: bool = True) -> None:
        od_kwargs = {} if od_kwargs is None else od_kwargs
        cl_kwargs = {"num_classes": 3} if cl_kwargs is None else cl_kwargs
        super().__init__(process_fn=process_data,
                         od_model=ObjectDetector, od_dataset_fn=SolitonODDataset,
                         od_loss_fn=MetzLoss, cl_model=SolDetClassifier,
                         cl_dataset_fn=SolitonClassDataset, cl_loss_fn=NLLLoss,
                         cl_aug=augment, od_aug=augment,
                         od_kwargs=od_kwargs, cl_kwargs=cl_kwargs,
                         stat_tools=[{"name": "pie classifier", "tool": PIEClassifier},
                                     {"name": "quality estimator", "tool": QE}],
                         stats_kwargs=[{"func": "modern", 
                                        "transformer": PowerTransformer},
                                       {"func": "modern",
                                        "transformer": PowerTransformer}])

        self.controllers["ML Controller"].get_tool("OD").metrics += [{"name": "Accuracy", "metric": od_accu_metric}]
        self.controllers["ML Controller"].get_tool("CL").metrics += [{"name": "Accuracy", "metric": cl_accu_metric}]
        self.controllers["Plot Controller"].add_new_tool(plot_tools=[{"name": "pie classifier", "tool": self.__plot_pie}])
        self.controllers["Plot Controller"].add_new_tool(plot_tools=[{"name": "quality estimator", "tool": self.__plot_qe}])

    def use_models(self, model_paths: list,
                   model_list: list | tuple = ("classifier", "object detector", "pie classifier", "quality estimator"),
                   data: list | dict | None = None) -> None:
        super().use_models(model_list=model_list, model_paths=model_paths, data=data)

        target_data = self.data if data is None else data
        for item in target_data:
            if "OD_pred" in item and type(item["OD_pred"]) is np.ndarray:
                item["OD_pred"] = pos_41labels_conversion(item["OD_pred"][0])
            if "CL_pred" in item and item["CL_pred"].flatten().shape[0] > 1:
                item["CL_pred"] = np.argmax(item["CL_pred"])
        if data is None:
            self.data = target_data
\end{lstlisting}
\end{figure}

A simplified overview of the detector configuration is shown in Listing~\ref{lst:soldet}.
The detector overrides the inherited \narrowtt{use\_models()} workflow in order to convert raw network outputs into application-specific predictions.
Classifier probability vectors are transformed into discrete excitation labels, while object-detector outputs are converted from compressed cell-space coordinates into image-space soliton positions.
The resulting predictions are stored directly in the detector data entries via the \narrowtt{CL\_pred} and \narrowtt{OD\_pred} fields.

\begin{figure}[!t]
    \centering
    \includegraphics[width=1\linewidth]{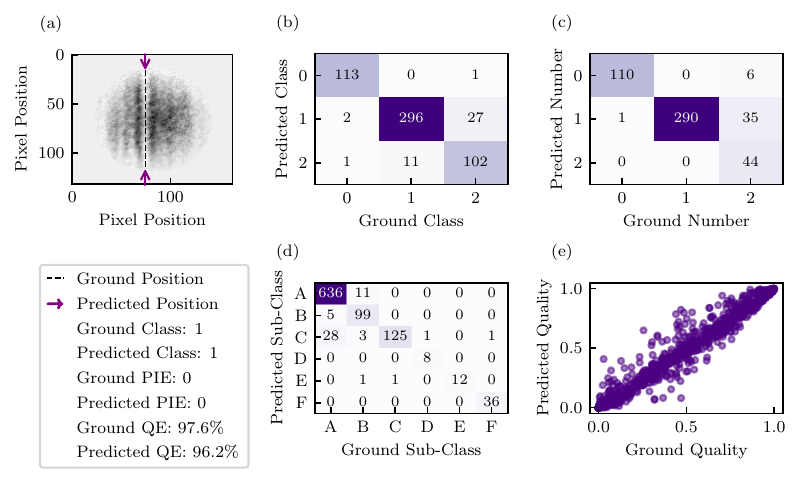}
    \caption{
    An overview of the performance of the modern SolDet sub-package.
    The ground-truth labels are taken from the \textit{Dark solitons in BECs dataset 2.0}~\cite{zwolak21_SDS}.
    (a) An example BEC from \narrowtt{class-1} with the object-detector position marked by purple arrows and the ground truth position shown in dashed black.
    The ground and predicted labels for the classifier, physically-informed excitation (PIE) classifier, and quality estimator (QE) are also shown in the legend.
    Confusion matrices for the (b) ML classifier and (c) object detector models that produced the highest $F_1$ scores on the ML test dataset. 
    (d) Confusion matrix for the PIE classifier and (e) a scatter plot for the QE.
    }
    \label{fig:soldet}
\end{figure}

Once initialized, the \narrowtt{SolitonDetector} can classify and track solitons in images of degenerate gases, such as the one shown in Fig.~\ref{fig:soldet}(a).
To evaluate the performance of the updated SolDet implementation, the public \textit{Dark solitons in BECs dataset 2.0}~\cite{zwolak21_SDS} was partitioned into training, validation, and test sets using the same procedures as in the original SolDet studies~\cite{guo21-MLD}.
The classifier was trained and tested using the \narrowtt{class-0}, \narrowtt{class-1}, \narrowtt{class-2}, and \narrowtt{class-8}\footnote{\narrowtt{class-8} is a subset of data originally assigned by human annotators to either \narrowtt{class-0}, \narrowtt{class-1}, or \narrowtt{class-2} but flagged as potentially mislabeled during the dataset curation process~\cite{fritsch22-DSD}.} data.
The object detector was trained using only the \narrowtt{class-0} and \narrowtt{class-1} data due to the lack of reliable position labels for \narrowtt{class-2}, though \narrowtt{class-2} data was used during testing.
Both models employed a 10-fold cross-validation with a 90:10 training-validation split.

Prior to testing, the test set was further filtered based on the initial label agreement among human annotators to facilitate a consistent comparison of performance with the legacy version of SolDet.
The test results were used to calculate weighted $F_1$\footnote{The $F_1$ score is a measure of a model's ability to make predictions. It gives equal weight to a model's precision in making a correct prediction for those it provides and its recall in producing all possible correct predictions.} scores for both the classifier and the object detector.
For the latter, the $F_1$ score was calculated based on the agreement between the number of identified excitations and the class label.
Confusion matrices for the ML classifier and object detector models with the highest $F_1$ score are shown in Fig.~\ref{fig:soldet}(b) and Fig.~\ref{fig:soldet}(c), respectively.
Additionally, $R^2$ scores of the excitation positions for the object detector were also calculated for each fold.
To handle object detector failure modes, such as false positives (detecting a soliton without a corresponding ground-truth position) or false negatives (failing to detect a soliton at a known position), missing position labels were manually assigned a value of zero.

After approximately \num{500} training epochs, the SolDet models achieved close agreement with the human-generated labels.
The classifier produced a mean weighted $F_1$ score of $91(1)~\%$, with the best-performing fold reaching $92.3~\%$.
This represents a marginal improvement over the original SolDet classifier reported in Ref.~\cite{guo21-MLD}, which achieved a best-model $F_1$ score of $91.6~\%$ on the corresponding consensus-labeled test subset, though this increase in performance might be in part due to the random initializations during each round of cross-validation.

The object detector achieved a mean weighted $F_1$ score of $89.0(8)~\%$, with the best-performing fold reaching $90.2~\%$.
This same fold reached an $R^2$ score of $99.95\%$ for the \narrowtt{class-1} subset of test data.
As shown in Fig.~\ref{fig:soldet}(c), the object detector reliably identifies the absence or presence of single solitons in \narrowtt{class-0} and \narrowtt{class-1}.
Performance on \narrowtt{class-2} data remains more variable due to the presence of multiple excitations and ambiguous non-solitonic structures.
These complex excitation patterns motivate the downstream physics-informed analysis stages described in the following subsection.

\subsubsection{Physics-informed statistical analysis}
Beyond ML-based localization and classification, the \narrowtt{SolitonDetector} incorporates downstream physics-informed statistical analysis for characterizing identified excitations.
These tools operate on the outputs of the ML workflow and provide physically interpretable descriptions of detected solitonic excitations.

The SolDet implementation includes two statistical analysis tools: the physically-informed excitation (PIE) classifier and the quality estimator (QE).
The PIE classifier categorizes detected solitonic excitations into one of six physically-motivated excitation subclasses summarized in Table~\ref{tab:classes}, while the QE estimates the degree to which a detected excitation resembles a canonical longitudinal soliton.
Additional details regarding the underlying physically informed analysis methods are provided in Ref.~\cite{guo22-CFS}.

\begin{table}[!t]
    \centering
    \begin{tabular}{cl}
        \hline\hline
        \multicolumn{1}{c}{\textbf{PIE Label}} &
        \multicolumn{1}{c}{\textbf{Description}} \\
        \hline
        A & Longitudinal solitons.\\
        B & Top partial solitons.\\
        C & Bottom partial solitons.\\
        D & Clockwise vortex solitons.\\
        E & Counterclockwise vortex solitons.\\
        F & Canted solitons.\\
        \hline\hline
    \end{tabular}
    \caption{
    The SolDet PIE classification taxonomy introduced in Ref.~\cite{guo22-CFS} organizes the data from \narrowtt{class-1} into physically-meaningful subclasses based on the type of excitation present. 
    }
    \label{tab:classes}
\end{table}

Both analysis stages are implemented within \texttt{Q-GAIN} using the \narrowtt{stat\_tool} interface described in Section~\ref{ssec:StatsControl}.
Each tool exposes \narrowtt{fit()} and \narrowtt{transform()} methods that integrate the corresponding analysis routines into the shared detector workflow.
This allows the statistical-analysis stages to operate directly on detector data entries produced by the ML pipeline while remaining compatible with the reusable \texttt{Q-GAIN} infrastructure.

The physically informed analysis routines are based on fitting procedures that estimate parameters describing the structure and geometry of detected solitons.
The implementation supports multiple fitting strategies, including Gaussian approximations, the original SolDet fitting routine, and a fitting implementation designed to better support images containing multiple excitations.
The resulting fitting parameters can additionally be transformed using configurable statistical transformations, with the default implementation employing a Yeo-Johnson transformation to improve the Gaussianity of the parameter distributions; see lines 12--17 in Listing~\ref{lst:soldet}.

The \narrowtt{SolitonDetector} extends the inherited detector workflow by incorporating the PIE and QE tools directly into the default analysis pipeline.
Following ML-based classification and localization, the detector applies the corresponding statistical analysis routines and stores their outputs in the detector data entries via the fields \narrowtt{pie classifier\_pred} and \narrowtt{quality} \narrowtt{estimator\_pred}.
An example of the resulting combined workflow output is shown in the legend of Fig.~\ref{fig:soldet}, while representative PIE-classification and QE-analysis results are shown in Fig.~\ref{fig:soldet}(d) and Fig.~\ref{fig:soldet}(e), respectively.

To evaluate the physically informed analysis stages, a fitting subset was constructed using $70~\%$ of each PIE subclass (A--F) of \narrowtt{class-1} data; see Table~\ref{tab:classes}. 
The PIE classifier was fitted using all PIE subclasses, while the QE was fitted only using subclass A corresponding to longitudinal solitons.
The remaining $30~\%$ of \narrowtt{class-1} data was reserved for testing.

The PIE classifier achieved a weighted $F_1$ score of $94.8~\%$ relative to the labels provided in the public \textit{Dark solitons in BECs dataset 2.0}~\cite{zwolak21_SDS}, while the QE metric achieved an $R^2$ score of $95.7~\%$.
Representative results are shown in Fig.~\ref{fig:soldet}(d) and Fig.~\ref{fig:soldet}(e).
Comparison with the legacy SolDet outputs reveals modest shifts in both PIE subclass assignments and QE scores.
In the PIE classifier, small changes are observed between longitudinal and partial-soliton PIE subclasses.
Larger changes occur in the QE predictions, where some excitations shift toward higher or lower estimated soliton quality.
These differences arise primarily from improvements introduced in the modified fitting procedures, particularly with changes designed to better support images containing multiple excitations.
Because the physically informed analysis methods depend directly on the distributions of fitted soliton parameters, modifications to the fitting workflow naturally alter the resulting statistical classifications.

\subsection{Vortex detection in ring BECs}
\label{ssec:vdet}
The vortex detector sub-package illustrates how \texttt{Q-GAIN} can be extended to support new scientific-analysis applications.
Whereas the SolDet sub-package demonstrated the modernization and integration of an existing workflow, the vortex detector was developed directly within \texttt{Q-GAIN} as a specialized object-detection pipeline for identifying quantized vortices in time-of-flight images of ring BECs.
Only the application-specific components---including the preprocessing routine, ML model, and inference logic---required implementation, while data management, model training, deployment, and export are handled through the shared \texttt{Q-GAIN} infrastructure.
The detector's object-detection workflow and neural-network architecture are inspired by the vortex-localization approach introduced by Metz \textit{et al.}~\cite{metz21-DLV}.

\subsubsection{VorDet implementation and workflow}
The \narrowtt{VortexDetector} subclasses the base \narrowtt{Detector} interface and integrates custom preprocessing routines, dataset wrappers, ML models, and loss functions through the controller architecture provided by \texttt{Q-GAIN}. 
A simplified overview of the detector configuration is shown in Listing~\ref{lst:vortex}. 
As with the other applications presented in this work, development is largely limited to implementing application-specific analysis modules, while the underlying infrastructure for data management, model training, deployment, and export is provided by the framework.

\noindent
\begin{figure}[!ht]
\begin{lstlisting}[
caption={A simplified overview of implementing the \narrowtt{VortexDetector}.
Full documentation and code for the supporting modules are available in the GitHub repository~\cite{doris25_QGN}.
Setting up \narrowtt{VortexDetector} requires providing a module to import the new data, a neural network suitable for object detection, and the supporting modules required by the ML controller, including a Dataset-compatible class used to convert the data into tensors and augment it, and a loss function for training.
Similar to the \narrowtt{SolitonDetector} implementation, the {\tt use\_models()} method is overridden to convert the model's output.},
label={lst:vortex}]
class VortexDetector(Detector):
    def __init__(self, **kwargs: dict) -> None:
        super().__init__(process_fn=vortex_process_fn,
                         od_model=ObjectDetector2D,
                         od_dataset_fn=VortexODDataset,
                         od_loss_fn=MetzLoss2D,
                         od_aug=True,
                         **kwargs)
        idx = self.controllers["ML Controller"].get_id(name="OD")
        self.controllers["ML Controller"].tools[idx]["tool"].metrics += [{"name": "Accuracy", "metric": accu_metric}]

    def use_models(self, model_paths: list, data: list | dict | None = None) -> None:
        super().use_models(model_list=("object detector"),
                           model_paths=model_paths, data=data)
        target_data = self.data if data is None else data
        for item in target_data:
            if "OD_pred" in item:
                item["OD_pred"] = vortex_labels_to_data(item["OD_pred"][0], threshold=(0.5, 8))
        if data is None:
            self.data = target_data
\end{lstlisting}
\end{figure}

The organization of the vortex datasets differs from the default structure expected by \texttt{Q-GAIN}, requiring a custom preprocessing function, \narrowtt{vortex\_process\_fn}, to import and transform the raw data. 
During preprocessing, time-of-flight images are cropped, masked, and optionally rescaled, while available metadata and vortex-position labels are extracted and associated with each measurement. 
The imported samples are then organized into training and testing subsets using the detector's tagging infrastructure.

The detector's ML workflow consists of a convolutional object-detection network, a PyTorch-compatible dataset wrapper, and a custom localization loss function. 
The dataset wrapper converts detector data into tensors suitable for training and optionally applies data augmentation through Gaussian noise, random rotations, horizontal and vertical reflections, and random translations. 
The object detector is trained using the localization loss introduced in Ref.~\cite{metz21-DLV}, enabling simultaneous prediction of vortex presence and spatial location.

Like the \narrowtt{SolitonDetector}, the \narrowtt{VortexDetector} represents detections internally using a compressed cell-space representation. 
Consequently, the inherited \narrowtt{use\_models()} method is overridden to convert the network output into physical image coordinates after inference. 
The converted vortex positions are then stored directly within the detector data entries, allowing them to be processed by the remaining \texttt{Q-GAIN} workflow without additional application-specific code.

\subsubsection{Performance}
\begin{figure}[!t]
    \centering
    \includegraphics[width=1\linewidth]{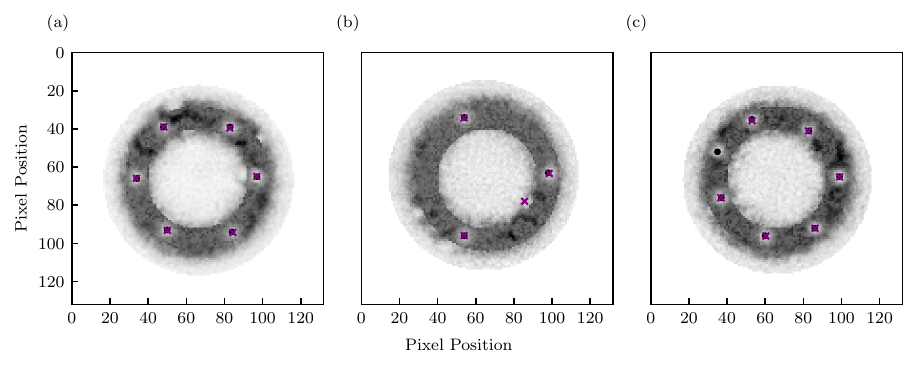}
    \caption{Example synthetic time-of-flight images of ring BECs with quantized vortices showing the ground truth (black circles) and predicted (purple ``x'') labels.
    (a) An example output for the object detector model with the highest $F_1$ score on the test set.
    For comparison, in (b) and (c), the model with the lowest $F_1$ score is used.
    (b) An example of a false-positive where the model has incorrectly identified a non-existent vortex.
    (c) An example of a false negative, where the model fails to identify a vortex.
    }
    \label{fig:vortex}
\end{figure}

The vortex detector is trained and evaluated using synthetic time-of-flight images of ring BECs, with representative examples from the test dataset shown in Fig.~\ref{fig:vortex}. 
Prior to training, the dataset is partitioned into 80:20 training and testing subsets. 
Model development employs 10-fold cross-validation on the training subset using a 90:10 training-validation split, with the final model selected based on the highest weighted $F_1$ score on the held-out test set.

The current VorDet implementation achieves a mean weighted $F_1$ score of $99.1(5)\,\%$ on the test dataset, with the best-performing fold reaching $100\,\%$.
Representative detector outputs are shown in Fig.~\ref{fig:vortex}(a).
Examination of the lower-performing models provides insight into the remaining failure modes.
False-positive detections occasionally occur when the network identifies ambiguous depletion features as vortices, as illustrated in Fig.~\ref{fig:vortex}(b).
Conversely, false-negative predictions arise when valid vortex signatures are missed, as shown in Fig.~\ref{fig:vortex}(c). 
These examples suggest that increasing the diversity of the training data, particularly by including more ambiguous depletion features, could further improve the detector's robustness.

Compared with the SolDet object detector, the current VorDet architecture is substantially larger, with \num{440340} trainable parameters versus \num{80306} for SolDet.
Consequently, a single \num{500}-epoch fold requires about one hour of training on an NVIDIA A100 GPU.

Although additional optimization is desirable to reduce model size and computational cost, the present implementation demonstrates that complex, application-specific analysis workflows can be developed in \texttt{Q-GAIN} using only a small number of detector-specific components.
Future work will focus on improving computational efficiency and extending the detector to more diverse experimental datasets and alternative trapping geometries.

\section{Conclusion}
\label{sec:conclusions}
We have presented \texttt{Q-GAIN}, a modular Python framework for constructing reusable ML and statistical-analysis workflows for scientific imaging applications.
The framework provides shared infrastructure for data management, preprocessing, ML training and inference, downstream statistical analysis, visualization, and export, while allowing application-specific analysis logic to be incorporated through extensible \narrowtt{Detector} and \narrowtt{Controller} interfaces.
By separating workflow infrastructure from domain-specific analysis modules, \texttt{Q-GAIN} enables scientific-analysis pipelines to be adapted, extended, and reused across a broad range of applications.

The examples presented in this work demonstrate complementary aspects of the framework architecture.
The MNIST workflow illustrates how existing ML models and statistical analysis tools can be integrated into \texttt{Q-GAIN} via lightweight application-specific wrappers.
The SolDet sub-package demonstrates how a mature scientific-analysis workflow can be incorporated into the framework while preserving domain-specific functionality and reusing shared infrastructure.
This updated revision of the SolDet implementation provides a PyTorch-based ML pipeline, greater flexibility in handling heterogeneous experimental data formats, and enhanced support for multiple excitations across the physically informed statistical analysis stages.
Finally, the vortex detector demonstrates how new scientific-analysis workflows can be developed within the framework using only a small number of application-specific modules layered on top of the reusable \texttt{Q-GAIN} infrastructure.

Together, these examples illustrate how \texttt{Q-GAIN} supports layered scientific-analysis workflows in which ML-based feature identification can be combined with downstream physics-informed statistical analysis within a unified execution environment.
The public release of the framework provides a foundation for continued development of new detectors, ML models, statistical-analysis tools, and scientific workflows through the \texttt{Q-GAIN} GitHub organization~\cite{doris25_QGN}.
As additional applications mature, they can be incorporated into the framework as reusable sub-packages while continuing to share common infrastructure for reproducible scientific analysis.

\section*{Acknowledgments}

\paragraph{Author contributions}
M. Doris: Software, Conceptualization, Validation, Methodology, Writing - original draft.
S. Guo: Software, Conceptualization, Investigation.
S.M. Koh: Software, Formal Analysis, Methodology.
L. Ritter: Software, Methodology, Conceptualization.
A.R. Fritsch: Software, Data curation, Investigation.
S. Mukherjee: Data curation, Investigation.
I.B. Spielman: Project administration, Resources, Writing - review \& editing.
J.P. Zwolak: Project administration, Resources, Software, Writing - review \& editing.

\paragraph{Funding information}
This work was partially supported by the National Institute of Standards and Technology; the Quantum Leap Challenge Institute for Robust Quantum Simulation (OMA-2120757); and the Air Force Office of Scientific Research Multidisciplinary University Research Initiative ``RAPSYDY in Q'' (FA9550-22-1-0339).

\paragraph{Note}
The views, conclusions, and recommendations contained in this paper are those of the authors and are not necessarily endorsed nor should they be interpreted as representing the official policies, either expressed or implied, of the U.S. Government. 
The U.S. Government is authorized to reproduce and distribute reprints for Government purposes notwithstanding any copyright noted herein. 
Any mention of commercial products is for information only; it does not imply recommendation or endorsement by the National Institute of Standards and Technology.


\end{document}